\documentclass[a4paper,11pt]{article}

\usepackage{jheppub} 
\usepackage[T1]{fontenc} 

\title{\boldmath Chiral Magnetic Effect in the Anisotropic Quark-Gluon Plasma}

\author[a,b]{Mohammad Ali-Akbari,}
\author[b]{Seyed Farid Taghavi}

\affiliation[a]{Department of Physics, Shahid Beheshti University, G.C., Evin, Tehran 19839, Iran}
\affiliation[b]{School of Particles and Accelerators, Institute for Research in Fundamental Sciences (IPM),
P.O.Box 19395-5531, Tehran, Iran}

\emailAdd{aliakbari@theory.ipm.ac.ir}
\emailAdd{s.f.taghavi@ipm.ir}

\newcommand{\bse}{\begin{subequations}}
\newcommand{\ese}{\end{subequations}}
\newcommand{\be}{\begin{equation}}
\newcommand{\ee}{\end{equation}}
\newcommand{\bea}{\begin{eqnarray}}
\newcommand{\eea}{\end{eqnarray}}
\newcommand{\ba}{\begin{array}}
\newcommand{\ea}{\end{array}}
\newcommand{\nn}{{\nonumber}}
\newcommand{\h}{\frac{1}{2}}

\def\FF{{\mathcal{F}}}
\def\HH{{\mathcal{H}}}
\def\B{{\mathcal{B}}}

\abstract{An anisotropic thermal plasma phase of a strongly coupled gauge theory can be holographically modelled by an anisotropic AdS black hole. The temperature and anisotropy parameter of the AdS black hole background of interest \cite{Mateos:2011tv} is specified by the location of the horizon and the value of the Dilaton field at the horizon. Interestingly, for the first time, we obtain two functions for the values of the horizon and Dilaton field in terms of the temperature and anisotropy parameter. Then by introducing a number of spinning probe D7-branes in the anisotropic background, we compute the value of the chiral magnetic effect (CME). We observe that in the isotropic and anisotropic plasma the value of the CME is equal for the massless quarks. However, at fixed temperature, raising the anisotropy in the system will increase the value of the CME for the massive quarks. }

\begin{document} 
\maketitle
\flushbottom

\section{Introduction and Result}
A new phase of Quantum Chromodynamics, called Quark-Gluon
Plasma (QGP) is produced at Relativistic Heavy Ion Collider (RHIC)
and Large Hadron Collider (LHC) by colliding two pancakes of heavy
nuclei such as Gold or Lead at a relativistic speed. From the numerical simulation, it is
realized that the QGP is a strongly coupled fluid since it has very low
viscosity over entropy density ratio. Therefore, perturbative calculations(methods) of QGP properties are not reliable \cite{CasalderreySolana}. Thus non-perturbative methods like AdS/CFT correspondence \cite{ads/cft} may be applied to describe different properties of the QGP such as rapid thermalization, elliptic flow,
jet quenching parameter and quarkonium dissociation which they have been considerably
studied in the literature \cite{Ollitrault,Snellings,CasalderreySolana}. The property we would like to discuss here is Chiral Magnetic Effect (CME) \cite{Kharzeev:2004ey, Kharzeev:2007tn, Kharzeev:2007jp,
Fukushima:2008xe,Kharzeev:2009fn}.

The presence of a strong magnetic field at the very early stages of heavy ion
collision, realized from analytical calculations \cite{Kharzeev:2007jp} and numerical simulation \cite{Voronyuk:2011jd}, 
and its accompanying non-trivial gluon field
configurations lead to the CME. More precisely, the axial charge $\mu_5$, given by the difference between the
number of fermions with left-handed and right-handed quarks, is
proportional to the  of non-trivial gauge field
provided that the left-handed and right-handed quarks are
initially equal.  The spin of quarks is tightly aligned along the
strong magnetic field. For a non-zero winding number, in order to
have a non-zero axial charge, the momentum direction of some quarks,
depending on the sign of winding number must be altered. This phenomenon produces 
a non-zero electric current of (massless) quarks along the strong magnetic field, which is given by %
\be\label{CME1} %
 J=\frac{\mu_5}{2\pi^2}B~. %
\ee %

Around one fermi after the collision, the produced plasma is thermalized and the system is in local equilibrium and hydrodynamics can be applied to study the evolution of the produced matter. At ultra-relativistic energies, the underlying physics enjoys axial global symmetry in the classical level, however, due to anomaly, this symmetry is broken in the quantum level. As a result, the anomalous hydrodynamic need to be considered \cite{Son:2009tf}. In this case, CME is emerged naturally by imposing the second law of thermodynamics. The ability of experimental observation of the anomaly effect in heavy ion experiment is still under intense debates \cite{Burnier:2011bf,Taghavi:2013ena,Belmont:2014lta}. Furthermore, the time dependent magnetic field in heavy ion collision is strong for $\tau \lesssim 1\,$fm and it becomes small afterward. The strong regime of magnetic field has overlap with a stage of plasma evolution that the system is not in local equilibrium  or in other words is not thermalized. In addition, at some stage of thermalization, the highly anisotropic energy-momentum tensor becomes isotropic and diagonal in the local rest frame of the plasma \cite{Chesler:2009cy}. In order to shed some light on the dynamic of CME in the early stages of heavy ion collision, we investigate the effect of simplified anisotropic environment, where the pressure in one direction of space is different from the other directions. We will use the AdS/CFT correspondence which is a powerful tool to study CME in the anisotropic strongly coupled QGP.



The AdS/CFT correspondence \cite{ads/cft} states that type IIB
string theory on $AdS_5\times S^5$ geometry, describing the near
horizon geometry of a stack of $N_c$ extremal D3-branes, is dual to
the four-dimensional ${\cal{N}}=4$ super Yang-Milles (SYM) theory
with gauge group $SU(N_c)$. In
particular, in the large $N_c$ and t'Hooft coupling limits, a strongly coupled SYM theory is
dual to the type IIb supergravity which provides a useful tool
to study the strongly coupled regime of the SYM theory. As a generalization,
a thermal SYM theory corresponds to the supergravity on an
AdS-Schwarzschild background where SYM theory temperature is
identified with the Hawking temperature of AdS black hole
\cite{Witten:1998zw}. Furthermore, Mateos and Trancanelli have introduced 
an interesting generalization of this duality to
the thermal and spatially anisotropic SYM \cite{Mateos:2011tv}.
In order to add matter (quark) in the fundamental representation of the corresponding gauge group, one needs to introduce a D-brane into the background in the probe limit \cite{Karch:2002sh}. The probe limit means that D-brane does not back-react on the geometry. Then the asymptotic shape of the brane gives the mass and condensation of the quark. In addition, the shape of the brane can be classified into two types, one is the Mikowski embedding (ME) and the other is black hole embedding (BE). While the ME does not see the horizon, the BE crosses it (for more details see appendix A).

To specify the anisotropic solution in \cite{Mateos:2011tv}, one needs to enter two inputs, the location of the horizon $u_h$ and the value of the Dilaton field at the horizon $\tilde{\phi}_h$. Then it is claimed that there is a one-to-one correspondence between these parameters and the anisotropy parameter $a$ and temperature of the system. It means that for given $u_h$ and $\tilde{\phi}_h$, using the equations of motion one finds just a temperature and an anisotropy parameter for the solution, though the map between ($u_h, \tilde{\phi}_h$) and $(a, T)$ has not been defined. In section 3, we carefully investigate this map and, for the first time, we find fitted functions between these parameters that is consistent with the numerical results. More precisely, we find the inverse map, meaning that, for given values of anisotrpy parameter and temperature one can find the corresponding $u_h$ and $\tilde{\phi}_h$. In fact this is one of the main results of this paper.

Applying the AdS/CFT correspondence, an interesting setup has been introduced in \cite{Hoyos:2011us} to describe the CME, as we will review and generalize it in the section 4. We realize that for the massless quarks the anisotropy of system does not affect the value of the CME and its value is the same as the isotropic case, \textit{i.e.} \eqref{CME1}. However, for the quarks with finite mass raise in anisotropy of the system will increase the value of the CME. In order to have non-zero value for the CME, the mass of the quark must vary between zero and its maximum at which this value vanishes. The maximim value of the mass also increases as one raises the anisotropy parameter. 

\section{Review on the Anisotropic Background}
The background we are interested in is an anisotropic solution of
the IIb supergravity equations of motion \cite{Mateos:2011tv}. This
solution in the string frame is given by %
\be\begin{split} \label{one}%
 ds^2&=g_{tt}dt^2+g_{xx}(dx^2+dy^2)+g_{zz}dz^2+g_{uu}du^2+g_{55} ds_{S^5}^2,\cr
ds_{S^5}^2&=d\theta^2+\cos^2\theta ds_{S^3}^2+\sin^2\theta d\psi^2,\cr
 \chi&=az,\ \ \phi=\phi(u).
\end{split}\ee %
$\chi$ and $\phi$ are axion and dilaton fields
respectively. $a$, which is a dimensionful constant, represents the anisotropy in the background. The components of metric are %
\be\begin{split}\label{two2}%
 g_{tt}&=-\FF\B u^{-2},\ g_{xx}=u^{-2},\ g_{zz}=\HH u^{-2}, g_{uu}=\FF^{-1}u^{-2},\ g_{55}=e^{\h\phi}.
\end{split}\ee %
$\HH$, $\FF$ and $\B$ depend only on the radial direction $u$.
In terms of the dilaton field, they have the following forms %
\bse\label{three}\begin{align} %
\HH&=e^{-\phi},\\
 \label{ff} \FF&=  \frac{e^{-\frac{1}{2}\phi}}{4(\phi'+u\phi'')}\left[ a^2 e^{\frac{7}{2}\phi}(4u+u^2\phi')+16\phi'\right]\,,
 \\
 \frac{\B'}{\B}&=\frac{1}{24+10
 u\phi'}\left(24\phi'-9u\phi'^2+20u\phi''\right)\,,
\end{align}\ese %
and the dilaton filed must satisfy an equation of order three
as %
\be\begin{split}\label{eomphi}%
 0&=\frac{256 \phi ' \phi ''-16 \phi '^3
   \left(7 u \phi '+32\right)}{u \, a ^2 e^{\frac{7 \phi }{2}}
   \left(u \phi '+4\right)+16 \phi '} +\frac{\phi ' }{u \left(5 u \phi '+12\right) \left(u \phi
   ''+\phi '\right)}\cr
   & \times \Big[13 u^3 \phi '^4+8 u \left(11
   u^2 \phi ''^2-60\phi''-12 u \phi ''' \right)
   +u^2 \phi '^3
   \left(13 u^2 \phi ''+96\right)  \cr
   & +2 u \phi '^2 \left(-5 u^3 \phi
   '''+53 u^2 \phi ''+36\right)+\phi ' \left(30 u^4 \phi
   ''^2-64 u^3 \phi '''-288+32 u^2 \phi
   ''\right) \Big] \,.
\end{split}\ee %
In the above equation, $a$ appears explicitly. One can shift the Dilaton field,
\be\label{tildephi}\begin{split} %
 \tilde{\phi}(u)=\phi(u)+\frac{4}{7}\log a,
\end{split}\ee %
to eliminate $a$.
Note also that the solution contains a self dual five-form
field(see appendix B for explicit form).
The horizon is located at $u=u_h$ meaning that $\FF(u_h)=0$ and the
Hawking temperature is given by %
\be %
 T=\frac{-\FF'(u_h)\sqrt{\B(u_h)}}{4\pi}. %
\ee %
The boundary lies at $u=0$ and the
metric approaches $AdS_5\times S^5$
asymptotically. At the boundary, the suitable boundary conditions are %
\bse\begin{align}%
 \label{BC1}\phi_B&=0,\\
 \label{BC2}\FF_B&=\B_B=1.
\end{align}\ese %

The gauge theory lives in a space-time with coordinates $(t,x,y,z)$.
Since there is a $U(1)$ symmetry in the $xy$-plane, $x$ and $y$ are
normally considered as the transverse directions and the
longitudinal direction is $z$. An anisotropy is clearly seen between
the transverse and longitudinal directions. For more details, we refer the reader to the original paper \cite{Mateos:2011tv}. The study of various properties of the above background has been done in the literature, e.g. see \cite{Ali-Akbari:2014xea}.

%

\section{More on the Anisotropic Background}

In order to study the CME as well as other physical quantities in the anisotropic background, one needs to specify the state of the medium under consideration by choosing appropriate values for the temperature $T$ and anisotropy parameter $a$. However, these parameters are related numerically to $u_h$ and the value of the deformed Dilaton field, introduced in \eqref{tildephi}, at the horizon, $\tilde{\phi}_h$. Since the inverse relation between $(u_h,\tilde{\phi}_h)$ and $(a,T)$ is not introduced in \cite{Mateos:2011tv}, specific values for $a$ and $T$ should be found by try and error. Moreover, the other difficulty is that, to achieve large values of $a/T$ in the field theory side, we need to fine-tune the values of $\tilde{\phi}_h$ and $u_h$. For example, choosing $(\tilde{\phi}_h=0.28, u_h=1)$ leads to $(a\simeq 508.135,T\simeq 0.692)$ and choosing $(\tilde{\phi}_h=0.28023, u_h=1)$ gives rise to $(a\simeq 6130.78,T\simeq 0.988)$. This means that for $\Delta \tilde{\phi}_h = 0.00023$, the difference between two $a/T$s is of the order of $5500$. Finally, the numerical calculation collapses. For instanse,  $\tilde{\phi}_h\gtrsim 0.281, u_h=1$. As we will show in appendix C, there is a curve 
\be \label{infinitAoT}
\tilde{\phi}_h(u_h)=\frac{2}{7}\log \frac{16}{6u_h^2},
\ee
where for an arbitrary point on this curve, $a/T$ goes to infinity. 

To bring the initial inputs of the anisotropic plasma more under control, we will study the inverse map between $\tilde{\phi}_h(a,T)$ and $u_h(a,T)$ in this section. Using Pad\'{e} approximation, we explicitly introduce a function to compute $\tilde{\phi}_h$ and $u_h$ in terms of $a$ and $T$. It should be noted that the method we will use here is based on numerical observation and is not a concrete analytical derivation. However, we will show different numerical evidences that the approximated inverse map is acceptable in a wide range of $a/T$.


\subsection{Map Between $(\tilde{\phi}_h, u_h)$ and $(a, T)$ }
As far as the authors of \cite{Mateos:2011tv} have been able to verify numerically, there is a one-to-one map between $(\tilde{\phi}_h,u_h)$ and $(a,T)$. As a matter of fact, for given values of $\tilde{\phi}_h$ and $u_h$, by solving the equations of motion \eqref{three}, $a$ and $T$ can be found. More explicitly, we have %
\bse \label{uhFiatemp} \begin{align}
 \label{uhFi} a(\tilde{\phi}_h,u_h)&=\lim_{\epsilon \to 0}\exp[\frac{7}{4}\tilde{\phi}(\epsilon;\tilde{\phi}_h,u_h)],\\
 \label{atemp}T(\tilde{\phi}_h,u_h)&= \sqrt{\B_h}\frac{a^{2/7} e^{-\frac{1}{2}\tilde{\phi}_h}}{16 \pi u_h}\left(16+u_h^2 e^{\frac{7}{2}\tilde{\phi}_h}\right),
\end{align}\ese %
where $\B_h=\B(\tilde{\phi}_h,u_h)$.
Equation \eqref{tildephi} and \eqref{BC1} indicate that the value of the anisotropy parameter is related to the asymptotic value of the $\tilde{\phi}(u)$, as it has been clearly emphasised in \eqref{uhFi}.

Now our goal is to investigate the inverse of mapping, if any, between $(\tilde{\phi}_h,u_h)$ and $(a,T)$. In order to do so, we start with equations of motion for the metric components \eqref{two2} (see equations (124-9) in \cite{Mateos:2011tv}). By introducing a new variable $u\rightarrow\xi u_h$, where $\xi \in [0,1]$ is a dimensionless variable, one can easily see that the equations of motion depend only on the variable $\xi$ and the dimensionless parameter $au_h$. As a result, this behavior points out that just the dimensionless parameter $au_h$ appears in the corresponding solutions of all metric components. 

By multiplying \eqref{atemp} by $1/a$, we find %
\be\label{Ta} %
 \frac{T}{a}=\sqrt{\B_h}\frac{e^{\frac{1}{2}\phi_h}}{16\pi a u_h}\left(16+a^2 u_h^2 e^{\frac{7}{2}\phi_h}\right).
\ee %
According to discussion presented in the previous paragraph, $\B_h$ and $\phi_h$ only turn out to be a function of the dimensionless parameter $au_h$ and therefore, from \eqref{Ta}, one can conclude that the same parameter appears in the $T/a$ or equivalently $T/a=f(au_h)$. In turn, it means that all the metric components are functions of $a/T$. Then it is easy to obtain 
\be\label{Tuh} %
  T u_h=\frac{1}{\eta}f^{-1}(\frac{1}{\eta}),
\ee %
where $\eta\equiv a/T$ and we have assumed that $f$ is invertible. 
\begin{figure}
\begin{center}
  \includegraphics[width=130mm]{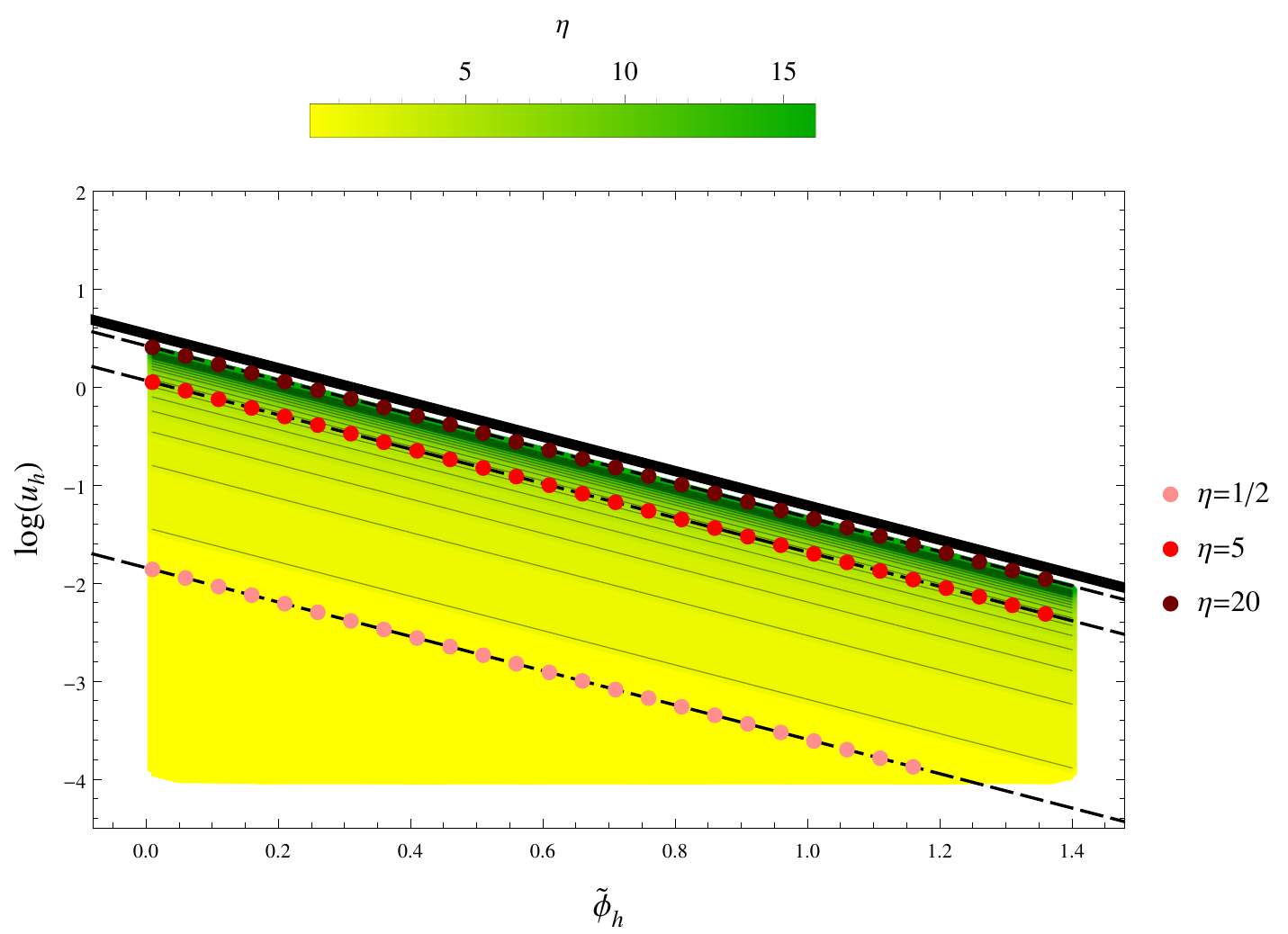}
\caption{ Contour plot of $\eta(\tilde{\phi}_h,\log u_h)$ is shown in the green region. Red dots indicate constant  lines of $\eta$ with accuracy $10^{-5}$. Black dashed lines are results of fit and thick black line corresponds to the case $\eta\to\infty$. }\label{separation}
\end{center}
\end{figure}

As a first step, we numerically plot $\eta(\tilde{\phi}_h,\log u_h)$. The plot has been shown in figure \ref{separation} by a green contour plot. In this figure, \eqref{infinitAoT} has been shown by the thick black line which corresponds to $\eta \to \infty$ and separates the $(\tilde{\phi}_h,\log u_h)$ plane into two regions. The upper part of the plane in not allowed and the lower part leads to finite values for $\eta$. The contour plot indicates that $\eta$ is constant over lines parallel to \eqref{infinitAoT}. To be more accurate, we also numerically find the points in the $(\tilde{\phi}_h,\log u_h)$ plane where for them the corresponding $\eta$ has constant values $1/2,\, 5$ and $20$ with accuracy $10^{-5}$. Then the fitted lines for these three different values of $\eta$ are shown in the figure \eqref{infinitAoT} by black dashed lines. This figure shows that all points with a specific value of $\eta$ is fitted by a line with slope $\sim -1.75$ and different intercepts. It suggests that for a constant value of $\eta$ we have $\log\, u_h = c(\eta) -1.75\, \tilde{\phi}_h$. By recalling that all metric components are functions of $\eta$ and assuming a generic form  $\log\, u_h = c(\eta) +\log \, \mathcal{K}(\tilde{\phi}_h)$,
the numerical observations give rise to 
\be \label{uheta}
u_h(\eta, \tilde{\phi}_h)=\kappa_1(\eta)\mathcal{K}(\tilde{\phi}_h),
\ee
where $\mathcal{K}(\tilde{\phi}_h)$ and $\kappa_1(\eta)$ are unknown functions. Thus from the above equation and \eqref{Tuh} one can find
\be \label{tildeFi02}
\tilde{\phi}_h=\mathcal{K}^{-1}(\frac{u_h}{\kappa_1(\eta)})=\mathcal{K}^{-1}(\frac{\kappa_2(\eta)}{T}),
\ee
where 
\be \label{DifKappa2}
\kappa_2(\eta)=\frac{1}{\eta}\frac{f^{-1}(\eta)}{\kappa_1(\eta)}.
\ee
We argued that $\phi_h$ should be a function of $\eta$. Using this fact, the far right equality of $\tilde{\phi}_h$ in \eqref{tildeFi02} and \eqref{tildephi} restricts the function $\mathcal{K}^{-1}(x)$ to be $-\frac{4}{7}\log x $ or
\be \label{KappaExp}
\mathcal{K}(x)=e^{-\frac{7}{4}x}.
\ee
Note that the factor $\frac{7}{4}=1.75$ is numerically consistent with our former result which was produced for a constant value of $\eta$. Using above equation and \eqref{tildeFi02} the inverse map for $\tilde{\phi}_h(a,T)$ can be written as
\be  \label{newphiuh1}
\tilde{\phi}_h(a,T)=\frac{4}{7}\log(\frac{T}{\kappa_2(a/T)}),
\ee
and for $u_h(a,T)$ we find 
\be  \label{newphiuh2} 
u_h(a,T)=\frac{1}{T}\kappa_1(\eta)\kappa_2(\eta),
\ee
where \eqref{Tuh} and \eqref{DifKappa2} have been used.
Henceforth, our aim is to gain the appropriate functions for $\kappa_1(\eta)$ and $\kappa_2(\eta)$. Unfortunately it is analytically impossible, and as we will explain in the following, suitable fitted functions will be found using the numerical results and asymptotic behaviours. 

\begin{figure}
\begin{center}
  \includegraphics[width=75mm]{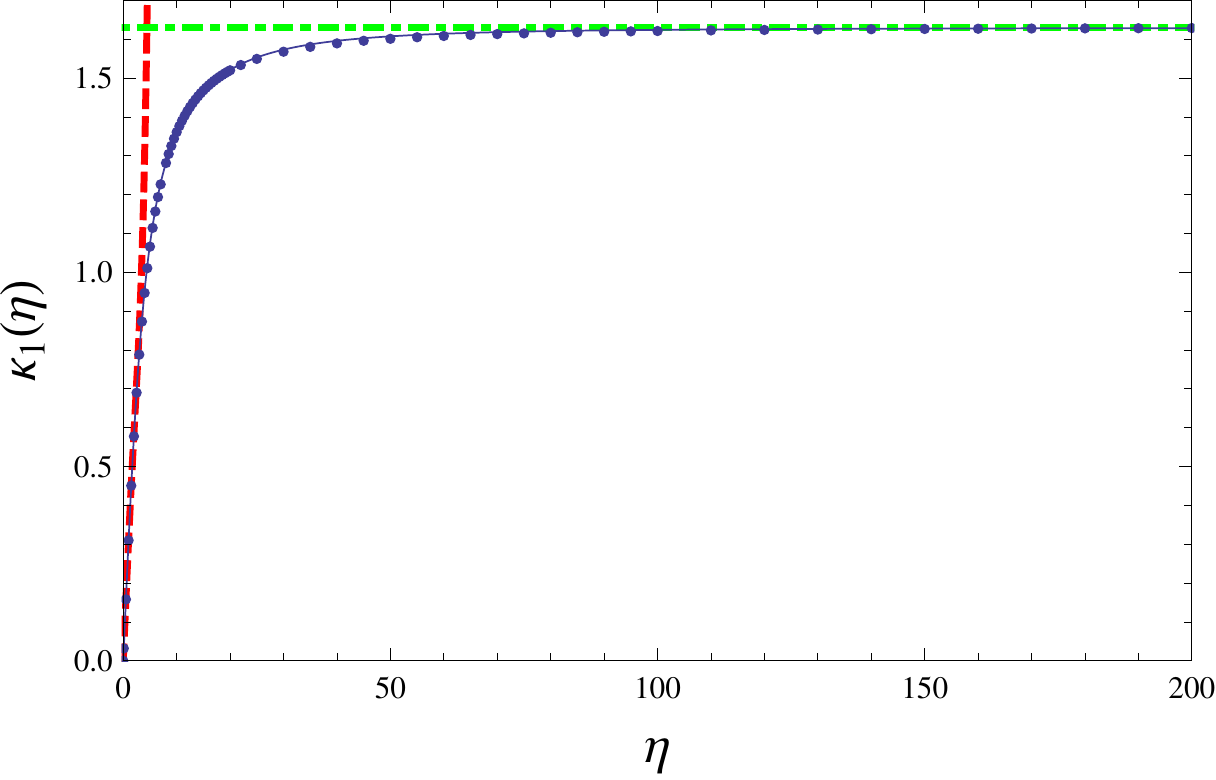}
  \includegraphics[width=75mm]{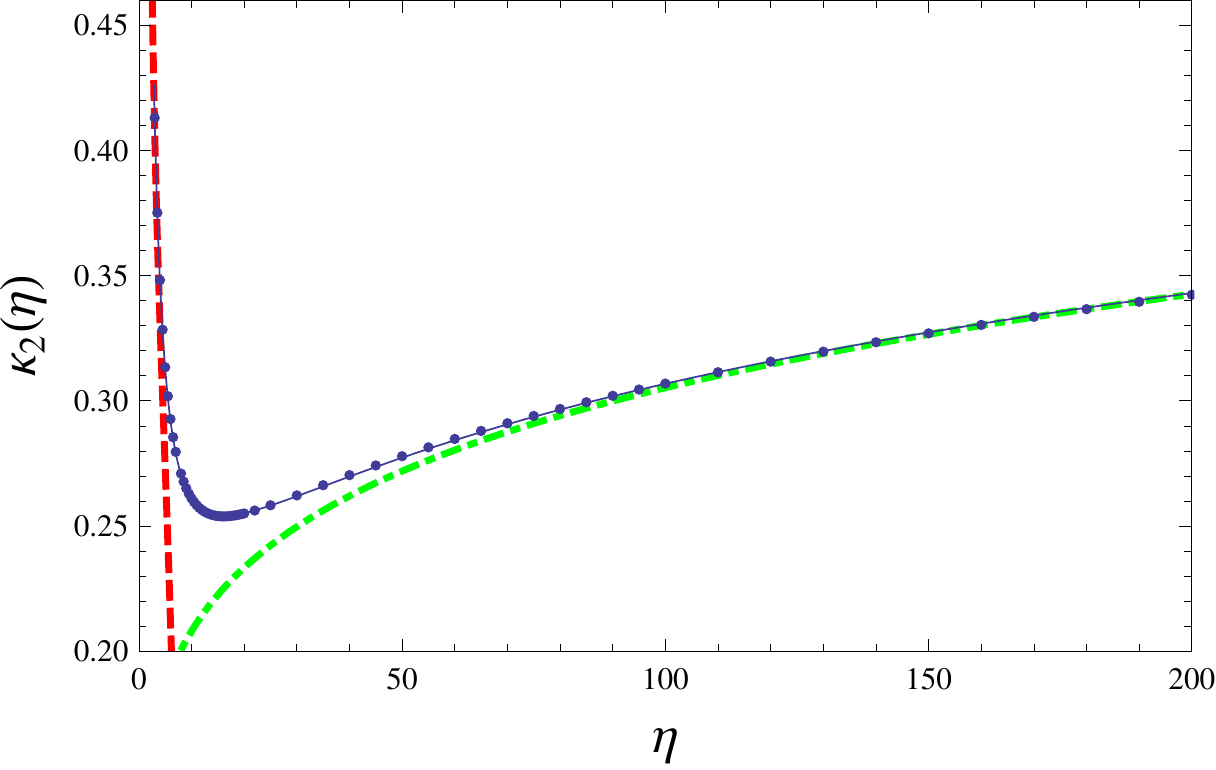}
\caption{ Numerical values for $\kappa_1(\eta)$ and $\kappa_2(\eta)$ (blue dots) and their analitical values for $T\gg a$ (red dashed curves) and $T\ll a$ (green dot-dashed curves). The blue curves represent the pad\'{e} functions \eqref{explicitform} for $\kappa_1(\eta)$ and $\kappa_2(\eta)$. }\label{kappa}
\end{center}
\end{figure}

\begin{itemize}

\item{\textbf{General values of $a$ and $T$}}

Now let us start with arbitrary values for $a$ and $T$, or equivalently $\eta$, and then solve the equation 
\be\begin{split} %
 \frac{a(\tilde{\phi}_h=0,u_{0h})}{T(\tilde{\phi}_h=0,u_{0h})}=\eta,
\end{split}\ee %
to find $u_{0h}(\eta)$. We can \textit{numerically} solve the above equation and according to \eqref{uheta} and \eqref{KappaExp} we have 
\be \label{kap1Num} \begin{split}  %
 \kappa_1(\eta)=u_{0h}(\eta).
\end{split}\ee%
From above equation and \eqref{newphiuh2}, it is straightforward to see 
\be \label{kap2Num}%
\kappa_2(\eta)=T(0,u_h=u_{0h}(\eta)).
\ee %
These functions have been shown in the figure \ref{kappa} by the blue dots. Notice that by knowing the functions $\kappa_1(\eta)$ and $\kappa_2(\eta)$ and using \eqref{newphiuh1} and \eqref{newphiuh2}, the inverse map can be found numerically. 

\item{\textbf{High temperature limit}}

In the case of high temperature, $T\gg a$, $\phi$ and $\pi T u_h$, up to $\mathcal{O}\left(\frac{\eta }{\pi }\right)^6$, can be written as \cite{Mateos:2011tv}
\bse\begin{align} 
  \label{farid1} \phi_h(\eta) &=-\frac{1}{4}\left(\frac{\eta}{\pi}\right)^2\left(\log 2 -\frac{\left(3-2 \pi ^2+72 (\log 2)^2\right) }{72 }\left(\frac{\eta}{\pi}\right)^2 \right),\\
 \label{farid2} \pi T u_h &=1+\left(\frac{\eta}{\pi}\right)^2\left(\frac{5\log 2 -2 }{48}+\frac{180+40\pi^2-12 \log 2 -273(\log 2 )^2}{13824}\left(\frac{\eta}{\pi}\right)^2 \right).
\end{align}\ese
From \eqref{tildephi} and \eqref{newphiuh1} we have
\be\label{ephih}
e^{\phi_h}= \left[\eta \kappa_2 (\eta)\right]^{-\frac{4}{7}},
\ee
and by considering \eqref{farid1}, it is easy to find that
\bea \label{K1}
\kappa_2(\eta)=\frac{1}{\eta }\left(1+\frac{7 \log 2 }{16 }\left(\frac{\eta }{\pi }\right)^2+\frac{7 \left(-12+8 \pi ^2-225 (\log 2)^2\right) }{4608 }\left(\frac{\eta }{\pi }\right)^4\right).
\eea
$\kappa_1(\eta)$ can be found using \eqref{newphiuh2}, \eqref{farid2} and \eqref{K1} and finally becomes
\bea \label{kapa2Asym}
\kappa_1(\eta) &=&\frac{\eta }{\pi }\left(1 - \frac{1+8 \log 2}{24 }\left(\frac{\eta }{\pi }\right)^2 \right.\nn\\
&+& \left.\frac{\left(108-32 \pi ^2+60 \log 2+1617 (\log 2)^2\right) }{3456 }\left(\frac{\eta }{\pi }\right)^4\right).
\eea
The red dashed curves in figure \ref{kappa}
represent the above functions.
 
\item{\textbf{Low temperature limit}}

In this limit the functions $\kappa_1(\eta)$ and $\kappa_2(\eta)$ can be gained in two ways. One way is to use the properties of the IR solution we have found in the appendix C. Another way is to use the numerical data from figure \ref{kappa}.\\
\textit{(i)}~ In the appendix C, a solution has been introduced in the low temperature limit, $a\gg T$. According to this solution, by comparing \eqref{newphiuh1} and \eqref{appendixphih} and utilizing \eqref{newphiuh2}, one gets %
\be\label{farid12} %
 \kappa_1(\eta)=\sqrt{\frac{8}{3}}\sim 1.63.
\ee %
Note that this is exactly the asymptotic value of $\kappa_1(\eta)$ in the limit of $\eta\rightarrow\infty$ in figure \ref{kappa}. Then using \eqref{newphiuh2}, \eqref{farid12} and \eqref{farid11}, one finds%
\be\label{khodam} %
 \kappa_2(\eta)=\frac{11^{7/12}}{4~ 2^{11/12} \pi^{7/6}}~\eta^{\frac{1}{6}}.
\ee %
\textit{(ii)}~ As it is clearly seen from figure \ref{kappa}, for large values of $\eta$ the function $\kappa_1(\eta)$ goes to a constant value, say $c\simeq1.63$. Thus \eqref{newphiuh1} and \eqref{newphiuh2} lead to 
\be\begin{split}
 \phi_h&=-\frac{4}{7}\log(\eta\kappa_2(\eta)),\cr
 u_h&=\frac{c}{T}\kappa_2(\eta).
\end{split}\ee 
Now the entropy density of the system can be computed via $s=\frac{N_c^2}{2\pi^2}\frac{e^{-\frac{5}{4}\phi_h}}{u_h^3}$ \cite{Mateos:2011tv} and above equations. In the end we have %
\be\label{entropy1}
 s=\frac{1}{2\pi c^3}N_c^2 T^3\eta^{\frac{5}{7}}[\kappa_2(\eta)]^{-\frac{16}{7}}.
\ee %
Furthermore, in this limit it was shown in \cite{Mateos:2011tv} that the entropy density scales as 
\be\label{entropy2} %
 s=c_{ent}N_c^2 a^{\frac{1}{3}} T^{\frac{8}{3}},
\ee %
where $c_{ent}\simeq 3.2$. Comparing \eqref{entropy1} and \eqref{entropy2}, one can identify $\kappa_2(\eta)$ with $(2\pi c^3 c_{ent})^{-7/16}\eta^{1/6}$ in agreement with \eqref{khodam}. 

Finally we have plotted the resulting functions for $\kappa_1(\eta)$ and $\kappa_2(\eta)$ (the green dot-dashed curves) in the figure \ref{kappa}. 
\end{itemize}

\subsection{Pad\'{e} approximant for $\kappa_1(\eta)$ and $\kappa_2(\eta)$}

We obtained the asymptotic forms of the $\kappa_1(\eta)$ and $\kappa_2(\eta)$ in the previous subsections in the region where $a \ll T$ and $a \gg T$. Applying the Pad\'e approximant and the mentioned asymptotic forms, one can gain explicit functions for $\kappa_1(\eta)$ and $\kappa_2(\eta)$ which fit the numerical results of the rest of the region. In order to be able to reproduce the asymptotic forms, we have to choose the following forms\footnote{Since, for large value of $a/T$($\gtrsim 30$),  $a$ and $T$ are too sensitive to initial inputs $\tilde{\phi}_h$ and $u_h$, the Pad\'e approximated functions \eqref{explicitform} do not reproduce the accurate results. Therefore one must use the numerical functions computed in \eqref{kap1Num} and \eqref{kap2Num} to find $\tilde{\phi}_h$ and $u_h$.}
\be\label{explicitform}\begin{split}
\kappa_1(\eta) &= \frac{\eta}{\pi}\left(\frac{1+\alpha_2 (\eta/\pi)^2+\alpha_4 (\eta/\pi)^4}{1+\beta_2 (\eta/\pi)^2+\beta_4 (\eta/\pi)^4+\beta_6 (\eta/\pi)^6}\right)^{\frac{1}{2}},\cr
\kappa_2(\eta) &=\frac{1}{\eta}\left(\frac{1+\tilde{\alpha}_2 (\eta/\pi)^2 + \tilde{\alpha}_4 (\eta/\pi)^4 +\tilde{\alpha}_6 (\eta/\pi)^6}{1+\tilde{\beta}_2 (\eta/\pi)^2 + \tilde{\beta}_4 (\eta/\pi)^4 }\right)^{\frac{7}{12}}.
\end{split}\ee
Moreover, for both functions we need to take into account the last terms in the numerator and denominator to have the best fits with the numerical results in the region with $T\sim a$ and as a result the parameters are
\be\label{para1}\begin{split}
\alpha_2 &\simeq 1.674349,\ \ \alpha_4 \simeq 0.076186,\ \ \beta_2 = \frac{1}{12} (1+12 \alpha_2 + 8 \log 2)  ,\cr 
\beta_4 &= \frac{1}{1728}\left(-99+144 \tilde{\alpha}_2 +1728 \tilde{\alpha}_4 +32 \pi ^2+84 \log 2+1152 \tilde{\alpha}_2 \log 2-1041 (\log 2)^2\right) ,\cr 
\beta_6 &= \frac{3}{8} \alpha_4 ,
\end{split}\ee
and %
\be\label{para2}\begin{split}
\tilde{\alpha}_2 &\simeq 0.927296,\ \ 
\tilde{\alpha}_4 \simeq  0.167441,\ \ 
\tilde{\beta}_2 = \tilde{\alpha}_2-\frac{3}{4} \log 2,\cr
\tilde{\beta}_4 &= \frac{1}{96} \left(3+96 \tilde{\alpha}_4 -2 \pi ^2-72 \tilde{\alpha}_2 \log 2 +99 (\log 2)^2\right),\cr
\tilde{\alpha}_6 &= \frac{11}{32} \tilde{\beta}_4
\end{split}\ee
Note that the three parameters $\alpha_6$, $\beta_2$ and $\beta_4$ in \eqref{para1} (and equivalently $\tilde{\beta}_2$, $\tilde{\beta}_4$ and $\tilde{\beta}_6$ in \eqref{para2}) are fixed by the asymptotic behaviour of $\kappa_1(\eta)$ and $\kappa_2(\eta)$ . It is important to notice that if we do not consider the last terms in the numerator and denominator of \eqref{explicitform}, all the free parameters can be fixed by the asymptotic forms of $\kappa_1(\eta)$ and $\kappa_2(\eta)$. In other words, one needs to have more terms in the expansion \eqref{K1} and \eqref{kapa2Asym}, for example up to $O(\frac{\eta}{\pi})^{10}$, to find all the free parameters, including $\alpha_6(\tilde{\alpha}_2)$ and $\beta_4(\tilde{\alpha}_4)$, by using the asymptotic behaviours.
\begin{figure}
\centering
  \includegraphics[width=6.9cm]{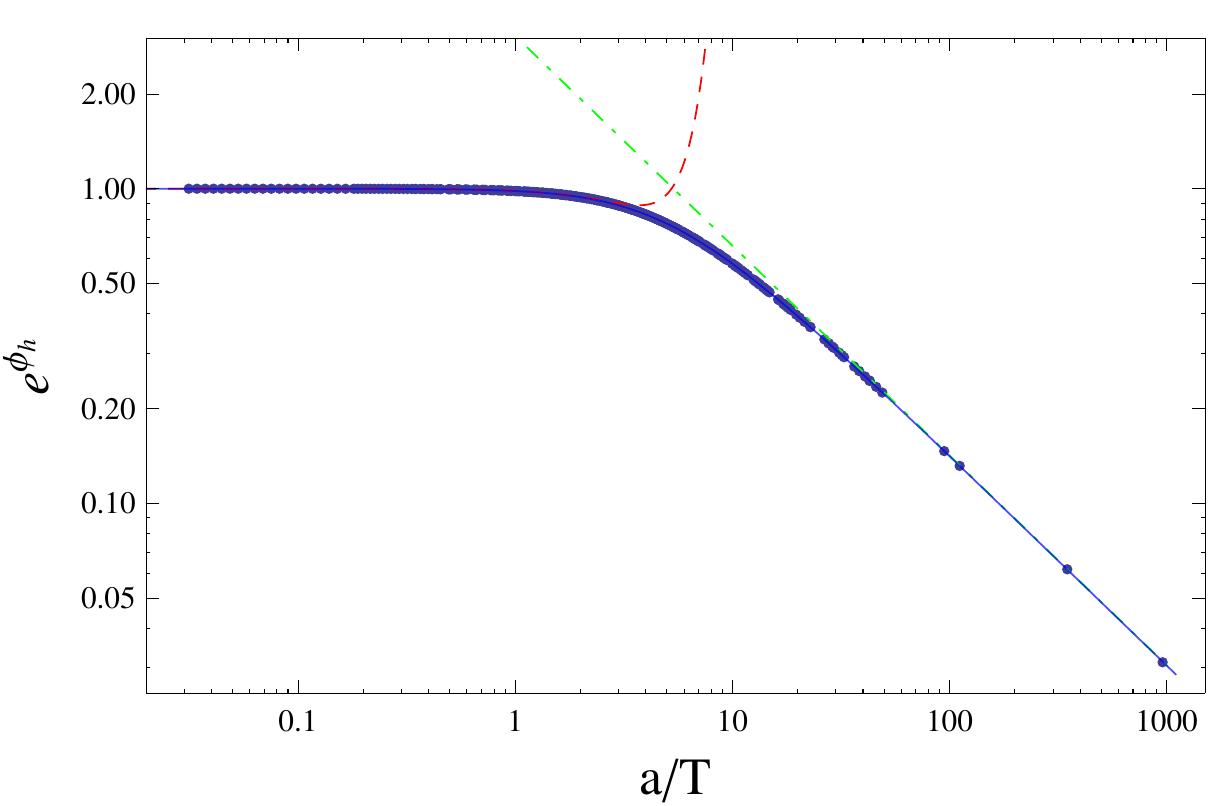}
  \includegraphics[width=6.9cm]{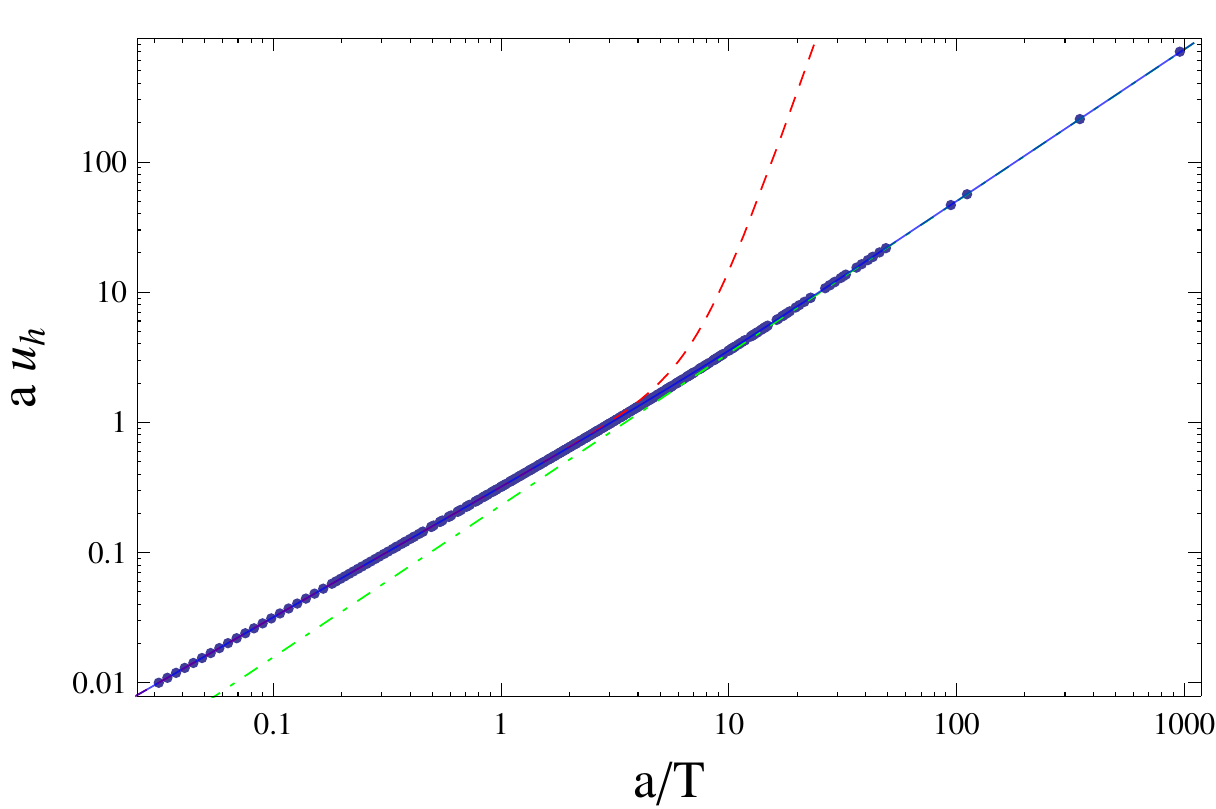}
\caption{ Numerical values for $e^{\phi_h}$ and $au_h$ (blue dots) and their corresponding high temperature (red dashed curves), low temperature (green dot dashed curves) and Pad\'e approximant (solid blue curves) versus $a/T$.}\label{fRatio}
\end{figure}

As a cross check and in order to clear more the validity of our assumptions, we will do the following numerical procedure, too. We solve the  equation of motion \eqref{eomphi} by choosing $\tilde{\phi}_h$ and $u_h$ as initial inputs. Then, using \eqref{uhFiatemp}, we find the values for $a(\tilde{\phi}_h,u_h)$ and $T(\tilde{\phi}_h,u_h)$ and \eqref{infinitAoT} gives the value of $\phi_h$. Now by choosing different values for $\tilde{\phi}_h$ and $u_h$, we plot $e^{\phi_h}$ and $a u_h$ with respect to $a/T$ which are shown by blue dots in figure \ref{fRatio}.
Note that these blue dots in figure \ref{fRatio} are  directly calculated numerically and neither the asymptotic analytical solutions nor separation of variable \eqref{uheta} has been used. In the other side, $e^{\phi_h}$ and $a u_h$ can be found by the method developed in this section. In figure \ref{fRatio}, we plotted $e^{\phi_h}$ and $a u_h$ using the analytical values of $\kappa_1(\eta)$ and $\kappa_2(\eta)$ for high and low temperatures and Pad\'e approximant with red dashed curve, green dot dashed curve and blue solid curve, respectively. It is not too hard to show that for a fix value of $a/T$ the ratio of numerically generated $e^{\phi_h}$ (or $a u_h$) to Pad\'e approximated function is less than $\sim 1.005$ which, in turn, indicates that our approximated functions work very well. On the other side, according to figure \ref{fRatio}, the Pad\'e approximated functions approach high temperature approximation values (red dashed curves) for $a/T \lesssim 1$ and low temperature approximation values (green dot dashed curves) for  $a/T\gtrsim 100$ which approve that our approximations seem to be valid in all the range of zero to infinity of $a/T$. 

Now by having the approximated functions of $\tilde{\phi}(a,T)$ and $u_h(a,T)$ at hand, the anisotropic metric for any desired values of $a$ and $T$ can be computed, easily. In the following we will use these functions and investigate the effects of anisotropic parameter and temperature on the CME.

\section{Holographic Setup of CME}
Since the QGP is a strongly coupled system, the AdS/CFT correspondence is a noticeable candidate to explain its properties.
Using the gravity dual, various properties of the plasma have been discussed. In particular, the CME has attracted much attention and an interesting gravity description of it has been introduced in \cite{Hoyos:2011us}\footnote{For another approach, \textit{e.g.} see \cite{Gahramanov:2012wz}.}. Such a description 
can be constructed of a supersymmetric intersection of $N_c$ D3-branes and $N_f$ rotating D7-branes as
\be\label{gauge} %
\begin{array}{ccccccccccc}
                   & t & x & y & z & u &  S^3 & \theta & \varphi  \\
                  D3 & \times & \times & \times & \times &  &  &  & \\
                  D7 & \times & \times & \times & \times & \times & \times &  &
\end{array}.
\ee %
Here D7-branes are rotating with angular velocity $\omega$ in $R\varphi$-plane  where $R=\sin\theta/u$. For more details on the coordinates of this plane we refer the reader to \cite{Hoyos:2011us}. The value of the angular velocity is identified with the axial chemical potential $\mu_5$ or more precisely $\omega=2\mu_5$\footnote{Consider ${\cal{N}} = 2$ SYM Lagrangian. After a chiral rotation  $\psi\rightarrow e^{-i\gamma^5\varphi/2}\psi$, the following new term appears in the fermion's kinetic term $$-\frac{\partial_\mu\varphi}{2}\bar{\psi}\gamma^\mu\gamma^5\psi. $$ Using $\varphi = \omega t$, it is evidently seen that $\omega = 2\mu_5$ (for more detail see \cite{Hoyos:2011us}).}. In the limit of large $N_c$ and large 't Hooft coupling constant $\lambda=g^2_{YM}N_c$,
the D3-branes are replaced by $AdS_5\times S^5$ background (they are replaced by AdS-Schwarzchild background at finite temperature). The system then reduces to $N_f$ rotating D7-branes in the AdS-Schwarzchild background with a worldvolume constant magnetic field which is needed to produce the CME. In the probe limit where $N_f\ll N_c$, the dynamics of the $N_f$ D7-branes on the AdS-Schwarzchild background, which is the gravitational dual of ${\cal{N}}=2$ SYM theory, is described by Driac-Born-Infeld (DBI) and Chern-Simons (CS) actions.

In the place of AdS-Schwarzchild background, let us start with a general background
\be\begin{split}\label{background}
ds^2=&-g_{tt}dt^2+g_{xx}(dx^2+dy^2)+g_{zz}dz^2+g_{uu}du^2 \cr
&+g_{ss} ds_{S^3}^2+g_{\theta\theta}d\theta^2+g_{\psi\psi} d\varphi^2,
\end{split}\ee %
which is asymptotically $AdS_5\times S^5$. $u$ is the radial coordinate with the boundary at $u\rightarrow 0$.
The ${\cal{N}}=2$ SYM theory lives in the Minkowski background with $t,x,y,z$. Moreover,  the above background contains a five-form field which asymptotically leads to the five-form field in the $AdS_5\times S^5$ background(for instance see appendix B).


In the low energy limit,  the action for the $N_f$ D7-branes in a general background is given by %
\be\begin{split}\label{action} %
 S &= S_{\rm{DBI}}+ S_{\rm{CS}}~,\cr
 S_{{\rm{DBI}}}&=-N_f \tau_{D7}\int d^8\xi\
 e^{-\phi}\sqrt{-\det(G_{ab}+2\pi\alpha'F_{ab})}~,\cr
 S_{\rm{CS}} &=N_f \tau_{D7}\int P[\Sigma C^{(n)}]e^{2\pi\alpha'F}~,
\end{split}\ee %
where $G_{ab}=g_{MN}\partial_a X^M\partial_b X^N$ is the induced metric on the probe branes. %
$\tau_{D7}^{-1}=(2\pi)^7l_s^8g_s$ is the D7-brane tension. $\xi^a$ are the
worldvolume coordinates and the capital indices $M,N,...$ are used
to denote space-time coordinates. In our case the background metric
$g_{MN}$ was introduced in \eqref{background}. $F_{ab}$ is the field
strength of the gauge fields living on the D7-branes.  As it was shown in \eqref{gauge}, the D7-branes extend along $t, x, y, z, S^3$ and the radial direction. In the CS
action, $C^{(n)}$ denotes Ramond-Ramond form fields and $P[...]$ is
the pull-back of the bulk fields to the worldvolume of D7-branes.

In order to describe the CME, we expect a current
caused by a magnetic field. We therefore consider
appropriate filed configurations on the D7-branes as follows \cite{Hoyos:2011us}
\be\begin{split} %
 \varphi(t,u)&=\omega t+\varphi(u),\ \ \theta(u),
\end{split}\ee %
and for the gauge field we consider the following cases
\bse\begin{align}%
 \label{z-direction} (i)~ A_z(u),\ \  F_{xy}=B_z,\\
 \label{y-direction} (ii)~ A_y(u),\ \  F_{xz}=B_y.
\end{align}\ese %
In the case $(i)$, the magnetic field is applied along the anisotropy direction. However, in $(ii)$ it is perpendicular to the anisotropy direction. Notice that metric translational symmetry in the $(x,y,z)$-directions and rotational symmetry in the 
$S^3$-directions allow us to choose above functionality for $\psi$ and $\theta$.
Using the AdS/CFT dictionary, dual operator coupled to $A_{z(y)}(u)$ is
$J^{z(y)}$ and the expectation value of $J^{z(y)}$ will describe the magnitude
of CME in the gauge theory side. $B_{z(y)}$ is a constant external
magnetic field. Here the axial chemical potential is described by
$\omega$. 
Substituting the above configurations in the action \eqref{action}, we find %
\be\begin{split} %
S_{DBI}&=-\int du \sqrt{Q_1+Q_2 A'^2_{z(y)}+Q_3\varphi'^2}, \cr
S_{CS}&=-\int Q_4 A'_{z(y)}.
\end{split}\ee %
Notice that $\prime=\frac{\partial}{\partial u}$ and %
\bse\begin{align}%
 Q_1&={\cal{Q}}_{z(y)}(g_{tt}-\omega^2 g_{\psi\psi})(g_{uu}+g_{\theta\theta}\theta'^2),\\
 Q_2&={\cal{Q}}_{z(y)}(g_{tt}-\omega^2 g_{\psi\psi})g^{uu},\\
 Q_3&={\cal{Q}}_{z(y)}g_{tt}g_{\psi\psi},\\
 Q_4&={\cal{N}}B\omega g_{ss}^2 ,
\end{align}\ese %
where 
\bse\begin{align} %
{\cal{Q}}_{z}&={\cal{N}}^2e^{-2\phi}g_{ss}^3 g_{xx}^2 g_{zz}(1+\frac{B_{z}^2}{g_{xx}^2}),\\
{\cal{Q}}_{y}&={\cal{N}}^2e^{-2\phi}g_{ss}^3 g_{xx}^2 g_{zz}(1+\frac{B_{y}^2}{g_{xx}g_{zz}}),\\
{\cal{N}}&= \frac{\lambda N_c N_f}{(2\pi^4)}.
\end{align}\ese %
Since the action depends only on the derivative of $A_{z(y)}$ and $\varphi$, there are two constants of motion, \textit{i.e.} $\alpha=\frac{\partial S}{\partial\varphi'}$ and $\beta=\frac{\partial S}{\partial A'_{z(y)}}$. After applying two successive Legendre-transformations with respect to $\varphi'$ and $A'_{z(y)}$, the final form of the action becomes 
\be %
 \hat{S}=-\int du \sqrt{\frac{Q_1}{Q_2}}\sqrt{Q_2(1-\frac{\alpha^2}{Q_3})-(\beta+Q_4)^2},
\ee %
where the hat means that the Legendre-transformations have been applied.
The location of the horizon on the probe branes, $u_*$, can be found by
\be\label{brane horizon} %
 Q_{2*}=Q_2(u=u_*)=0. 
\ee %
Then the reality condition on the action implies that \cite{Hoyos:2011us}
\bse\begin{align} %
 \label{reality1} \alpha&=-\sqrt{Q_{3*}}, \\
 \label{reality2} \beta&=-Q_{4*},
\end{align}\ese %
where $Q_3(u)$ and $Q_4(u)$ are evaluated at $u=u_*$. 

According to the gauge-gravity correspondence, the expectation value of the dual operators $J^{z(y)}$ and $O_\varphi$ coupled to $A_{z(y)}$ and $\varphi$ 
can be found from the asymptotic expansions of $A_{z(y)}$ and $\varphi$. It was shown in \cite{Hoyos:2011us} that 
\bse\begin{align} %
 \label{beta} \langle J^{z(y)} \rangle &=-(2\pi\alpha')\beta, \\
 \label{alpha} \langle O_\varphi \rangle &=\alpha.
\end{align}\ese %
As a result, $\beta$ up to a constant, gives the value of the CME on the gauge theory side. Also regarding the discussion about discrete space-time symmetries, $\alpha$ is an order parameter of spontaneous symmetry breaking \cite{Hoyos:2011us}. Furthermore, since the background \eqref{background} is asymptotically $AdS_5\times S^5$, the asymptotic expansion of $\theta$ is given by \cite{Hoyos:2011us, Karch:2005ms}
\be\label{mass} %
\theta(u)=\theta_0 u+\theta_3 u^3+...~.
\ee %
The leading term $\theta_0$ is proportional to the mass of the fundamental matter and $\langle O_m\rangle\propto \theta_3$ where $O_m$ is the operator dual to mass.

\begin{figure}
\begin{center}
  \includegraphics[width=61mm]{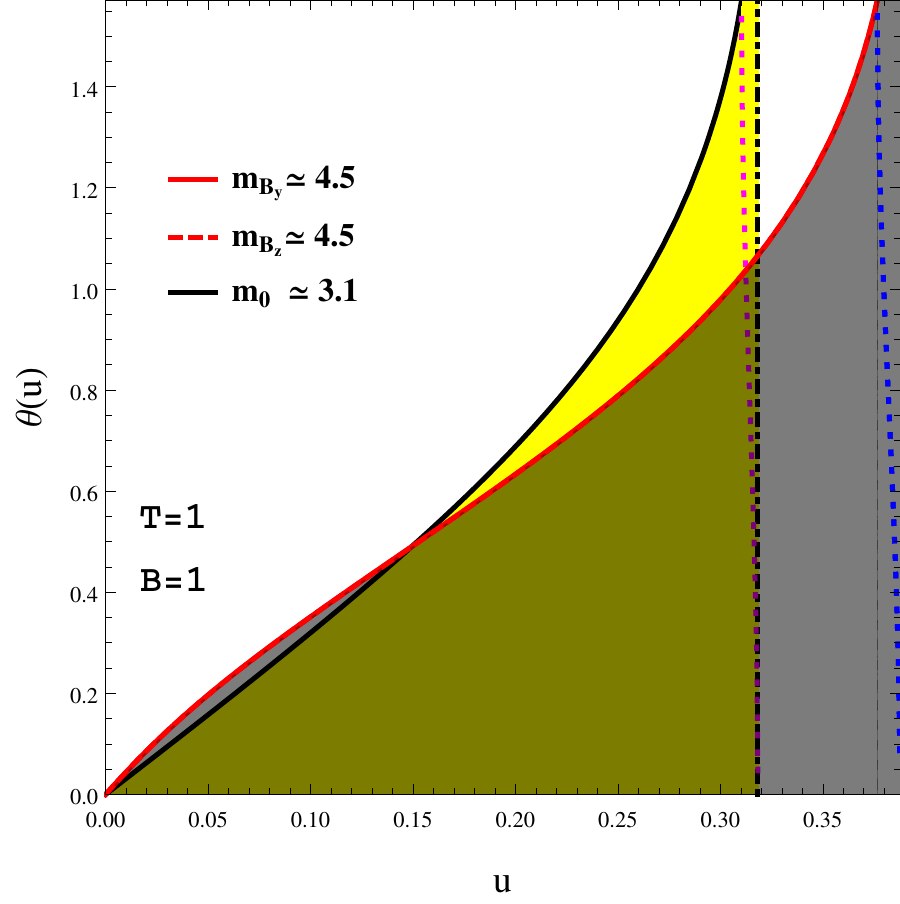}
  \includegraphics[width=61mm]{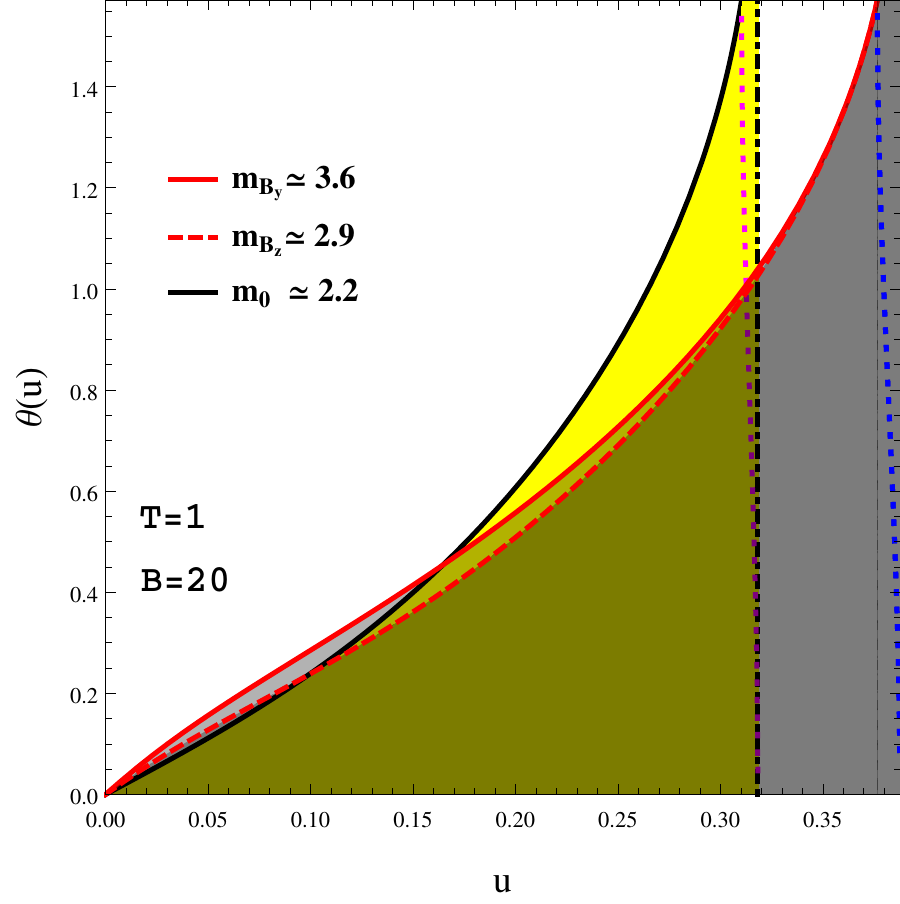}
  \includegraphics[width=61mm]{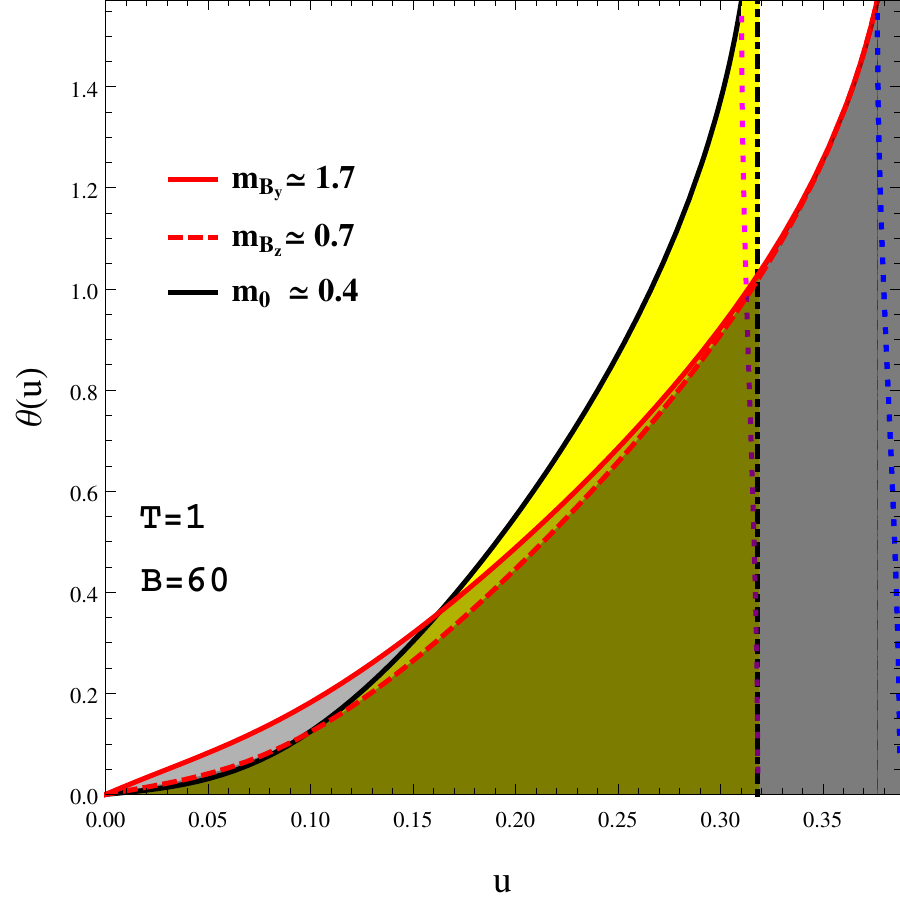}
  \includegraphics[width=61mm]{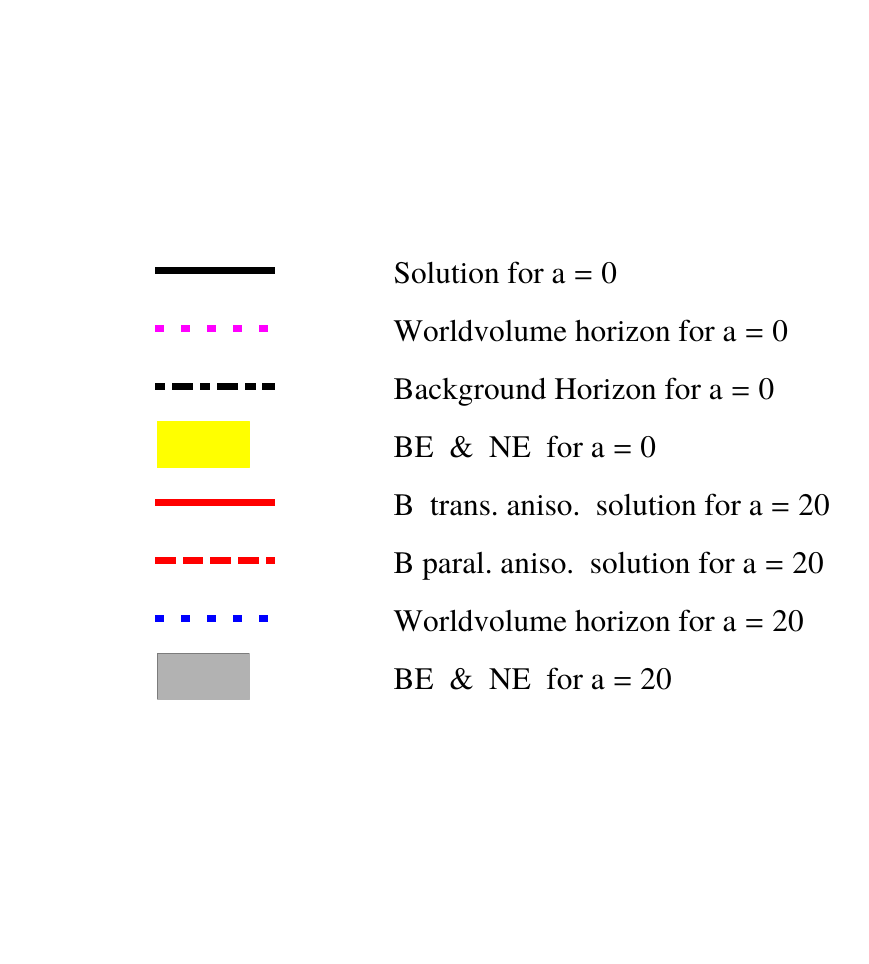}
\caption{ Various embeddings of the probe brane in the presence of the external magnetic field. The black dot-dashed line and the right axis illustrate the locations of background horzon for the cases $a=0$ and $a=20$, respectively. The worldvolume horizons have been indicated by a doted curve for $a=0$ and a blue curve for $a=20$. The yellow and gray regions correpond to the NEs and BEs. The black curve shows the critical embedding for $a=0$. The embedding corresponding to both the red (with magnetic field transverse to the anisotropic direction) and the red-dashed (with magnetic field parallel to the anisotropic direction) curves are the same.}\label{embedding}
\end{center}
\end{figure}

\section{Numerical Results for the CME}
Our goal in this section is to compute the value of the CME in terms of the quark mass in the anisotropic background \eqref{one}. To do so, we should solve the equation of motion for $\theta(u)$ to find the quark mass from \eqref{mass}. Since it is straightforward to solve this equation of motion, we do not mention it here and refer the reader to \cite{Hoyos:2011us, AliAkbari:2012if}. It is worth noticing that, in order to find the solutions, $\beta$ must be chosen on the brane horizon $u_*$. Therefore, apart from MEs, BEs and NEs(see appendix A) have non-zero current and contribute to the CME. In figure \ref{embedding} the right axis shows the location of the background horizon, $u_h\simeq 0.39$, for $T=1$ and $a=20$ and the background horizon at $u_{0h}=0.32$ has been shown by vertical dot-dashed line when $T=1$ and $a=0$. The solutions in the yellow and gray regions cross the worldvolume horizon and correspond to the NEs and BEs with non-zero currents.

\subsection{ Zero Mass Case}
Although in the presence of the magnetic field the trivial solution $\theta(u)=0$ is not a favourable solution energetically, let us start with this exceptional case. \eqref{mass} indicates that the mass of the fundamental matter is zero for this spacial case. Moreover, from \eqref{brane horizon}, it is easy to see that the horizon on the probe D-branes coincides with the horizon in the background, \textit{i.e.} $u_*=u_h$. As a result, according to \eqref{reality2} and \eqref{beta}, the value of the CME is 
\be %
 \langle J^{z(y)} \rangle= \langle J^{z(y)}_0 \rangle,
\ee %
where $ \langle J^{z(y)}_0 \rangle=(2\pi\alpha')^2{\cal{N}}\omega B_{z(y)}= \frac{N_c N_f}{2\pi^2}\mu_5B_{z(y)}$ is the value of the CME in the isotropic background (or equivalently in the isotropic SYM theory) \cite{Hoyos:2011us, AliAkbari:2012if}.  In other words, via a holographic calculation one realizes that the value of the CME is insensitive to the anisotropy of the system in the massless case.

\subsection{ Finite Mass Case }
For non-trivial solutions $\theta(u)\neq 0$, as it was explained in \eqref{mass}, the asymptotic value of $\theta(u)$, or more precisely $m\propto\lim_{\epsilon \to 0}\theta'(\epsilon)$, specifies the mass of the fundamental matter. The value of the mass generally depends not only on the magnetic field but also on the anisotropy parameter. In figure \ref{embedding}, we have plotted the critical embedding, horizons on the brane and in the background in terms of various values of the magnetic fields and two values of anisotropy parameter at a fixed temperature. Let us summarize the main points: 
\begin{itemize}
\item In the presence of the anisotropy parameter for small but equal values of $B_y$ and $B_z$, the value of the mass does not change. However, the difference between $m_{B_y}$ and $m_{B_z}$ is significant for larger values of the magnetic fields (see the red curves in the figure \ref{embedding}). Note that $m_{B_{y(z)}}$ denotes the mass of the quarks when the magnetic field is applied along $y(z)$. 
\item When a large magnetic field is applied along the anisotropy direction, the value of the mass can be smaller than it is in the isotropic case. Therefore, the masses for which the CME is non-zero is more restricted than the case with $a=0$. Notice that for the small value of the magnetic fields, the allowed masses are extended.
\item When the magnetic fields apply transverse to the anisotropy direction, the allowed masses with non-zero value of the CME are extended.
\end{itemize}
Some of the above results are in agreement with the ones obtained in \cite{Ali-Akbari:2013txa}.

\begin{figure}
\begin{center}
  \includegraphics[width=67mm]{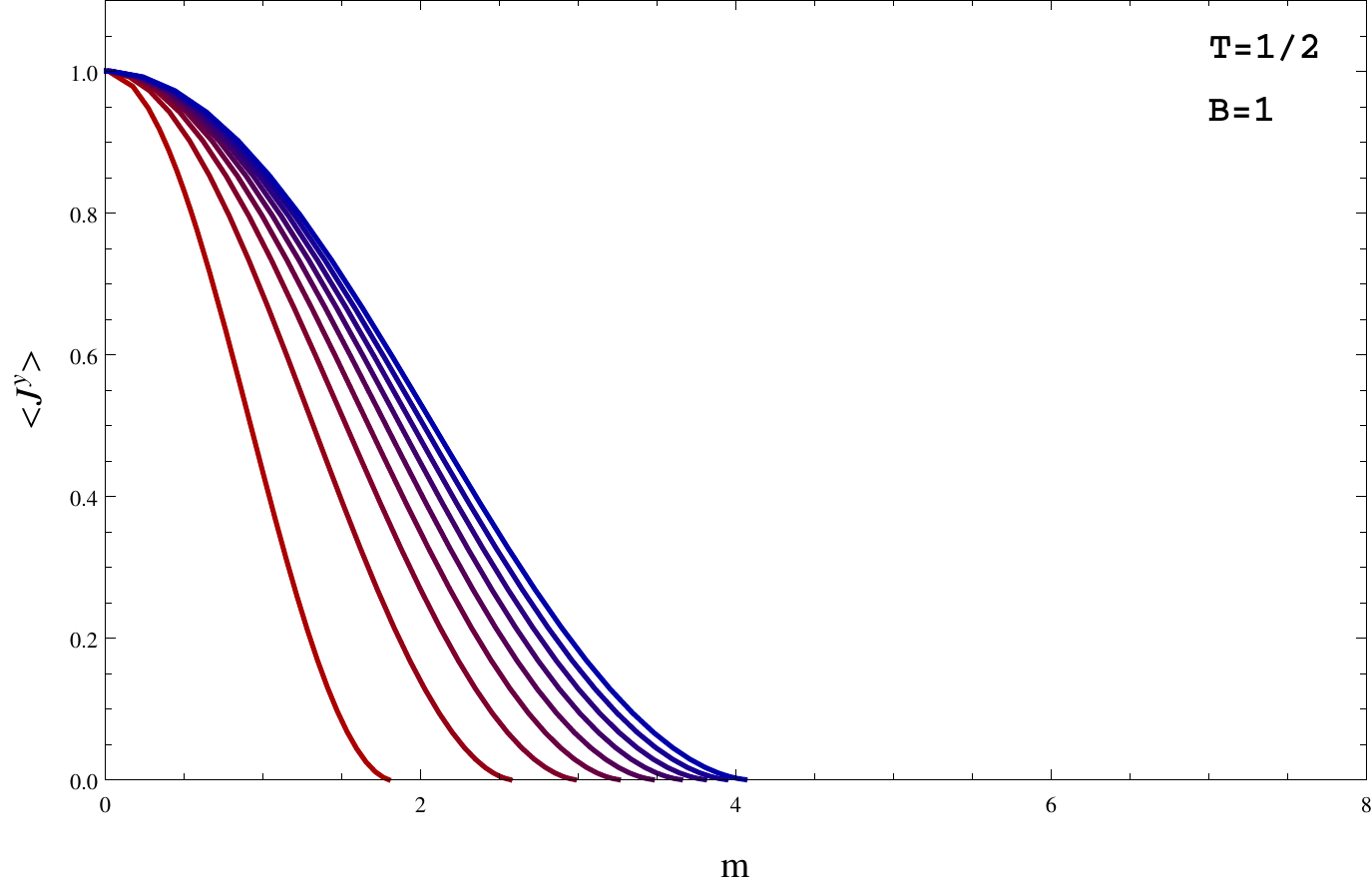}
    \includegraphics[width=81mm]{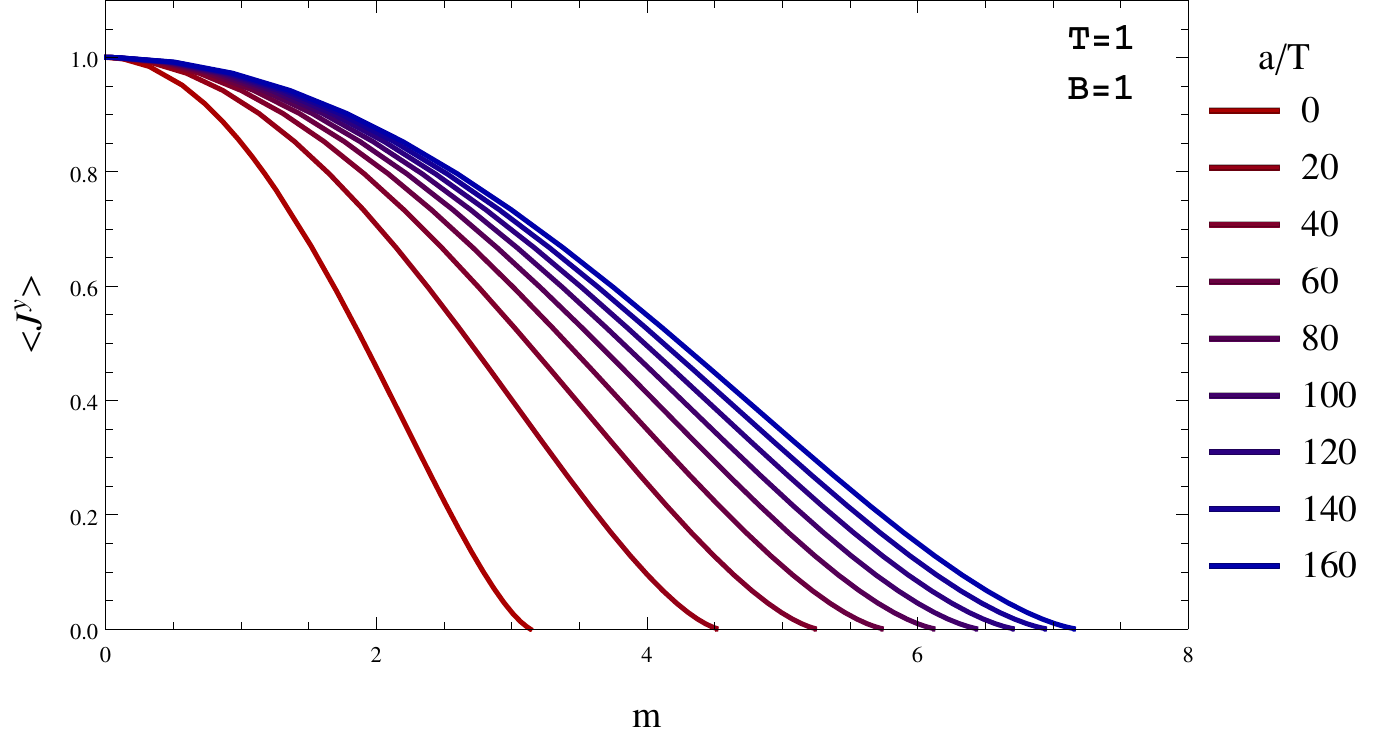}
\caption{ The value of the CME as a function of the quark mass for $B=1$ and $T=1/2$ (left) and $T=1$ (right) and different values of $a$. }\label{current}
\end{center}
\end{figure}

In \cite{Hoyos:2011us}, it was shown that raising the temperature in the system will increase the value of the CME. This behaviour also persists in the presence of the $\alpha'$-correction \cite{AliAkbari:2012if}. The numerical computation whose results are plotted in figure \ref{current} indicates that this value increases as one raises the anisotropy in the system at fixed temperature. As a matter of fact, the anisotropy parameter somehow behaves similarly to the temperature, in agreement with the (numerical) results displayed in the literature \cite{Ali-Akbari:2013txa, Chakraborty:2012dt}. Moreover, notice that there is critical mass at which the CME vanishes. Higher values of the mass have zero current. In this figure, although  we plotted $J^y$ versus mass, $J^z$ behaves similarly for $B=1$.


\section{Conclusion}
In this paper, we have studied the effects of the anisotropy parameter and the temperature on the CME. The effects are investigated for general values of both parameters by considering the rotating probe D7-branes in the anisotropic background. Our main findings can be summarized as follows. 
\begin{itemize}

\item Utilizing the asymptotic forms in the regions with $a\ll T$ and $a\gg T$, we have introduced the functions \eqref{explicitform} which can be used to find the values of the $u_h$ and $\tilde{\phi}_h$ for given values of $a$ and $T$; \textit{i.e.} $\tilde{\phi}_h(a,T)$ and $u_h(a,T)$. In fact this is the first time that this map has been introduced.

\item 
In the second part of the paper,  applying the functions \eqref{explicitform}, we observed that the mass, at which the CME vanishes, becomes larger with respect to the case where $a=0$ by raising the anisotropy parameter at fixed temperature.  

 
\end{itemize}

%
%

\appendix
\section{ Embeddings of the Probe Brane}

In the probe limit, the embeddings of a probe D-brane are classified into three categories according to its shape in the anisotropic background. As it was stated in the introduction, MEs are those embeddings that close off above the background horizon. In other words, there is no horizon on the probe branes. On the contrary, BE means that the probe brane sees the background horizon and its horizon is precisely coincident with the background one. It is well-known that the quark-antiquark bound states (mesons) are stable on the MEs. However, they are unstable on the BEs. In the presence of the electric field which is turned on the brane \cite{Nakamura:2012ae} or for rotating probe branes \cite{Ali-Akbari:2013tca}, a new group of embeddings (NEs) appears. As a matter of fact, for this group of solutions there is a horizon on the probe brane which is \textit{not} coincident with the background horizon. On
the MEs since the quark-antiquark bound states are stable, there
are no free charge carriers and consequently no current and hence the system behaves as an insulator. Oppositely, on the BEs and NEs the bound states are unstable and as a result,
non-zero current is observed \cite{Hoyos:2011us, Kim:2011qh}.

\begin{figure}
\begin{center}
  \includegraphics[width=61mm]{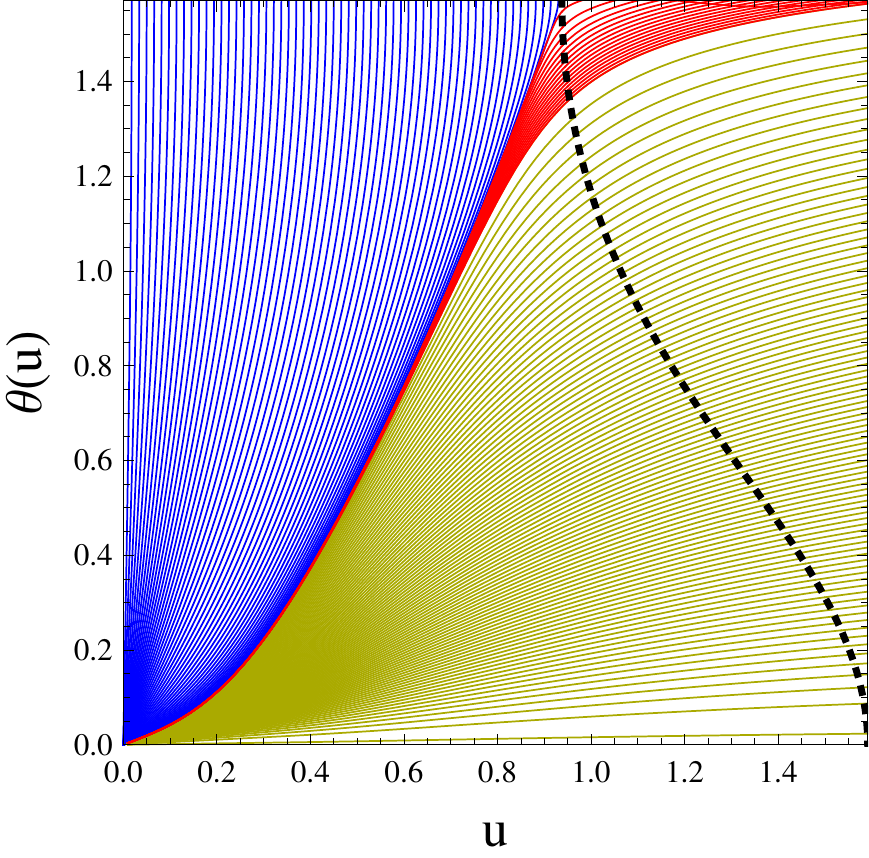}
\caption{ A plot of the possible D7-brane embeddings for $u_h=5/\pi, B=8$ and $\omega=1$.}\label{embedding1}
\end{center}
\end{figure}

A set of possible brane embeddings is presented in figure \ref{embedding1}. The background horizon is located at $u\sim 1.6$ and therefore the blue and green curves show MEs and BEs, respectively. The black dashed curve represents the horizon on the probe brane and since there is a cross point between red curves and this horizon, the red curves show NEs. Moreover note that NEs exist in a narrow region of mass.

\section{Five-Form Field Notation}
The five-form is taken to be proportional to the volume form of
the five-sphere
\be %
 F_5=\alpha(\Omega_5+\star\Omega_5)
\ee %
where $\star$ denotes Hodge star operator. Assuming that the metric of the five-sphere is  
\be %
 ds^2_5=d\theta_1^2+\sin^2\theta_1 d\phi_1+\cos^2\theta_1(d\gamma_1^2+\sin^2\gamma_1 d\gamma_2^2+\sin^2\gamma_1\sin\gamma_2 d\gamma_3^2).
\ee %
It is then straightforward to write the volume form of the five-sphere in terms of the above coordinates as  
\be\begin{split} %
 \Omega_5&=\cos^3\theta_1\sin\theta_1\sin^2\gamma_1\sin\gamma_2~ d\theta_1 \wedge d\phi_1 \wedge d\gamma_1 \wedge d\gamma_2 \wedge d\gamma_3, \cr
 \star\Omega_5&=-\frac{e^{-\frac{7}{4}\phi}}{u^5}\sqrt{\B}~ dt \wedge dx \wedge dy \wedge dz \wedge du,
\end{split}\ee %
and therefore the components of the five-form are given by 
\be\begin{split} %
 F_{txyzu}&=-\frac{\alpha e^{-\frac{7}{4}\phi}}{u^5}\sqrt{\B},\cr
 F_{\theta_1\phi_1\gamma_1\gamma_2\gamma_3}&=\alpha\cos^3\theta_1\sin\theta_1\sin^2\gamma_1\sin\gamma_2.
\end{split}\ee %
Since $F_5=dC_4$, we consider the following ansatz  
\be\begin{split} %
 C_4=C_{txyz} ~dt \wedge dx \wedge dy \wedge dz+ C_{\phi_1\gamma_1\gamma_2\gamma_3}~ d\phi_1 \wedge d\gamma_1 \wedge d\gamma_2 \wedge d\gamma_3,
\end{split}\ee %
and then one can simply find that 
\be\begin{split} %
 C_{txyz}&=-4\alpha\int\frac{du}{u^5}e^{-\frac{7}{4}\phi}\sqrt{\B} ,\cr
 C_{\phi_1\gamma_1\gamma_2\gamma_3}&=-\alpha\cos^4\theta\sin^2\gamma_1\sin\gamma_2.
\end{split}\ee %

\section{The IR solution}
For large values of the temperature, $T\gg a$, it is possible to find analytic expressions for
the metric and the dilaton \cite{Mateos:2011tv}. 
In this appendix we will discuss and analytically find an interesting solution of the equation of motion \eqref{eomphi} in the low temperature limit. To do so, let us start with the following ansatz 
\bea \label{IRdilatonSolution}
\phi(\xi) =\lim_{\varepsilon \to 0}\left( \phi_h-\frac{4}{7} \log \xi + \varepsilon f(\xi) \right),
\eea
where we assume that $f(\xi)$ and its derivative are finite at $\varepsilon \to 0$ and $\xi=\frac{u}{u_h}$. The explicit form of the $\phi_h$ can be found by using the consistency condition for the solution introduced in \cite{Mateos:2011tv} (see Equ. (139)) which is given by %
\be %
 \tilde{\phi}'_h=-\frac{4 e^{\frac{7}{2}\tilde{\phi}_h}u_h}{16+e^{\frac{7}{2}\tilde{\phi}_h}u_h^2}
\ee %
and we then have %
\be\label{appendixphih} %
 \tilde{\phi}_h=\frac{2}{7}\log\frac{16}{6u_h^2}
\ee %
Regarding above equation, by substituting ansatz \eqref{IRdilatonSolution} in the \eqref{eomphi} one can see that it satisfies the equation of motion provided that we choose %
\be %
 f(\xi) =c_1 \xi^{11/7-\sqrt{55/7}}+c_2 \xi^{11/7+\sqrt{55/7}}
\ee %
where $c_1$ and $c_2$ are arbitrary constants. Therefore, \eqref{IRdilatonSolution} leads to %
\be %
 \phi(u)= \frac{4}{7}\log \frac{\sqrt{8/3}}{au},
\ee %
and using \eqref{three} the other components of the metric can be found as below %
\bea
\HH(u) &=& \left(\frac{3}{8}\right)^{2/7}\left(a u\right)^{4/7}, \\
\FF(u) &=&  \frac{49}{11}\left(\frac{1}{18}\right)^{3/7} \left(a u\right)^{2/7},\label{IR_F} \\
\B(u) &=& \frac{11}{49}18^{3/7} \left(\frac{1}{a u}\right)^{2/7}.
\eea
\eqref{IR_F} reveals that the proposed solution describes a zero temperature background or more precisely $T\ll a$. As a result, this solution can be considered as an IR limit of a general anisotropic background. The temperature and entropy of the solution are then straightforward to be computed form \eqref{Ta} and $s=\frac{N_c^2}{2\pi^2}\frac{e^{-\frac{5}{4}\phi_h}}{u_h^3}$ \cite{Mateos:2011tv}. Hence we obtain 
\bea \label{IRTemp}
\label{farid11}\frac{T}{a} &=& \left(\frac{\sqrt{11}}{\pi 2^{17/14} 3^{3/7} }\right) \left(\frac{1}{a u_h}\right)^{6/7},\\
s &=& \frac{N_c}{2 \pi} \left(\frac{3}{8} \right)^{5/14} a^{5/7} u_h^{-16/7}.
\eea
Eliminating $u_h$ between two above equations, we find %
\bea
s = c_{ent} N_c^2 a^{1/3} T^{8/3}
\eea
where $c_{ent}= \frac{2^{7/6}3^{3/2}\pi ^{5/3}}{11^{4/3}}=3.2$, compatible with the numerical value for $c_{ent}$ in \cite{Mateos:2011tv}.

\section{Effect of the Magnetic Field on the Mass}
Using set of \eqref{gauge} coordinates, it is convenient to introduce the new following coordinates  
\be\begin{split} %
 R&= \frac{1}{u} \sin\theta, \cr
 r&= \frac{1}{u} \cos\theta.
\end{split}\ee %
In this coordinate system, the shape of the probe D7-branes is described by $R(r)$, as it was extensively studied in the literature and the probe branes are rotating in $R\phi$-plane with angular velocity $\omega$. Moreover the asymptotic value of $R(r)$ is identified with the mass of the quarks, \textit{i.e.} $m=\lim_{r \to \infty} R(r)$. 
\begin{figure}
\begin{center}
  \includegraphics[width=120mm]{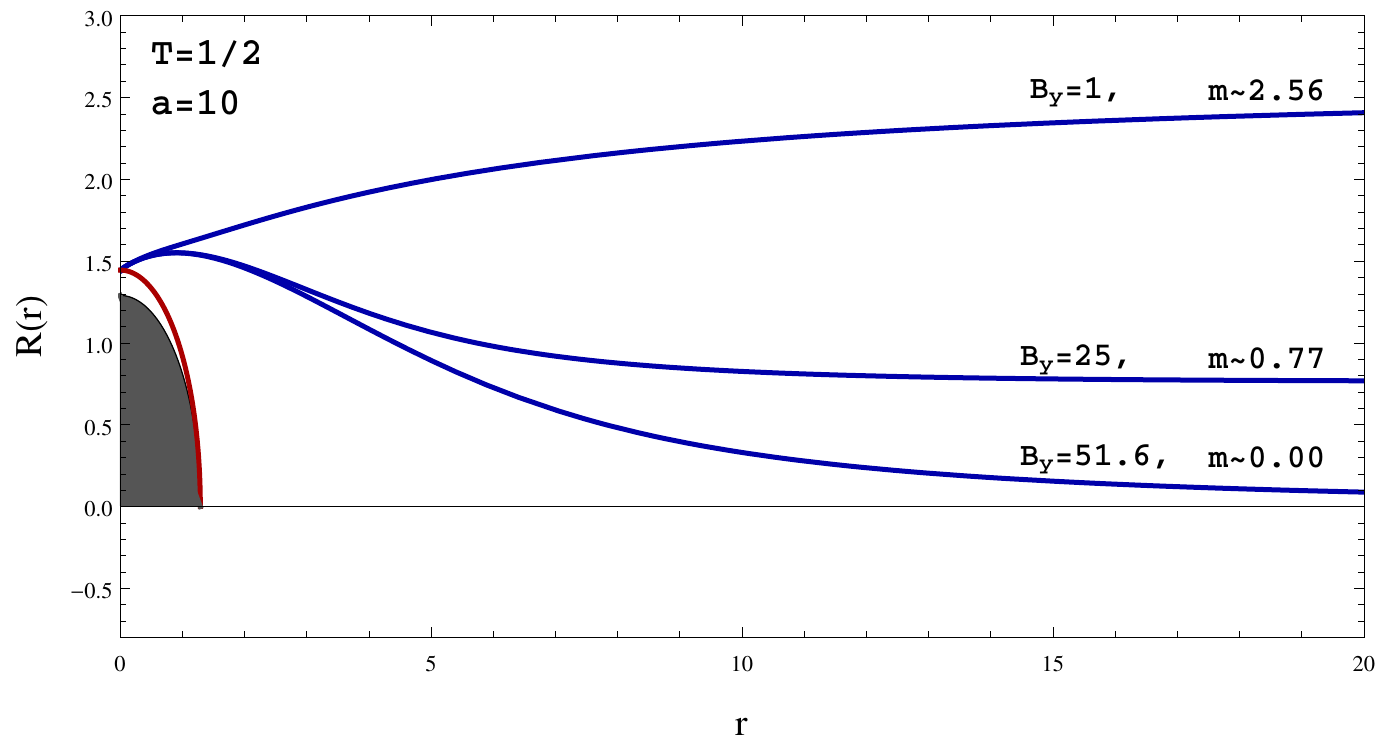}
\caption{ The critical embedding of probe D7-branes for $T=1/2$ and $a=10$ when magnetic field is perpendicular to the anisotropy.  
}\label{RrB}
\end{center}
\end{figure}
\begin{figure}
\begin{center}
  \includegraphics[width=75mm]{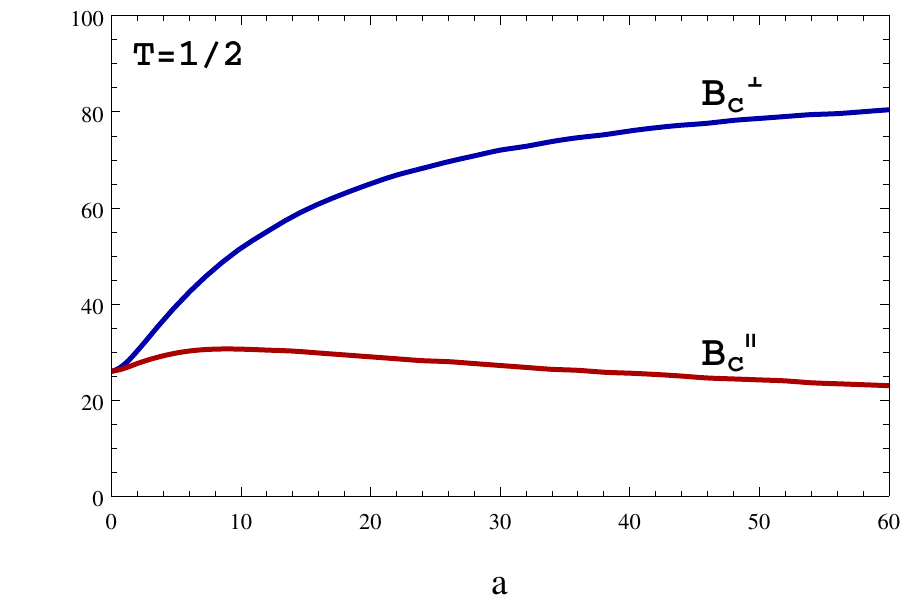}
  \includegraphics[width=75mm]{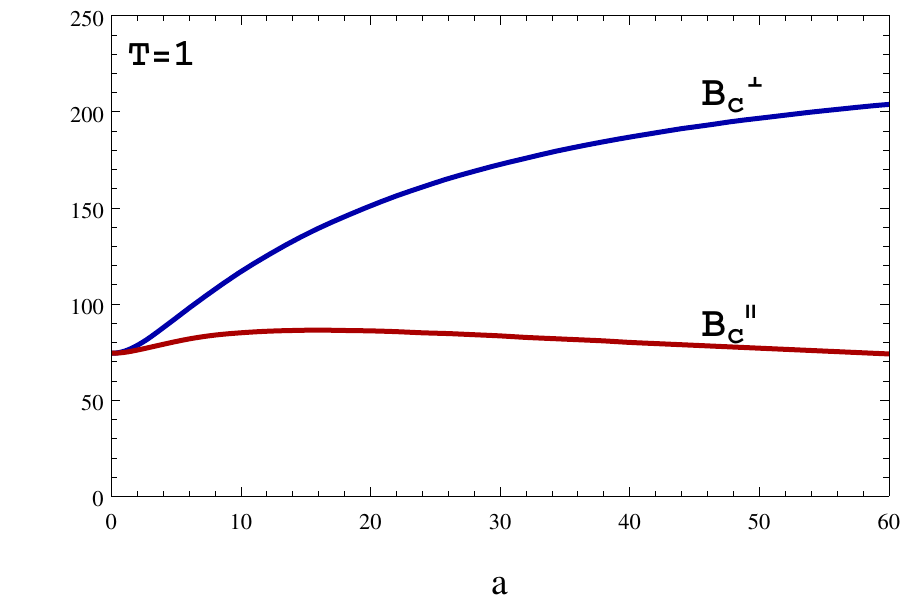}
\caption{The critical value of magnetic field as a function of anisotropy $a$ for two cases, magnetic field perpendicular and parallel to anisotropy. }\label{criticalMagnet}
\end{center}
\end{figure}
When the magnetic field is non-zero, as it is evidently seen from
figure \ref{RrB}, the value of the mass decreases as the magnetic
field increases. Therefore, for fixed values of the temperature and anisotropy parameter 
an upper limit exists for the magnetic field. For larger values of the
magnetic field the mass of the quark becomes negative and the
corresponding configurations are not physical \cite{Babington:2003vm}.

We finally investigate the regime of validity of the parameters $a$, $T$ and $B_{y(z)}$. The behaviour of D7-brane solutions in the presence of magnetic field and chemical potential has been extensively studied in the literature \cite{O'Bannon:2008bz, Ali-Akbari:2013txa, Filev:2007gb}. 
In fact, for given values of $a$ and $T$, it is well known that there is a critical value for the magnetic field $B_c$ at which MEs \footnote{MEs, NEs and BEs have been introduced in appendix A.} start to have a negative asymptotic value at $r\rightarrow \infty$, corresponding to the negative mass for the matter fields. Furthermore, when the magnetic field is larger than its critical value, one can check that for any given NE or BE(which has a non-zero current) there is a ME(indicating a zero current) with the same mass where its free energy is lower. It means that the later embedding is thermodynamically preferred and therefore no current is induced in the system.

In the presence of the anisotropy parameter, the magnetic field can be applied in the parallel or perpendicular to the anisotropy direction. In figure \ref{criticalMagnet}, we depict the critical values of magnetic field for $T=1/2$ and $T=1$. This figure illustrates that for the small values of anisotropy parameter, the critical magnetic field does not depend on the anisotropy,  drastically. But by increasing the anisotropy parameter, the difference between the values of the critical magnetic field is noticeable.

\end{document}